\documentclass[prl,aps,twocolumn,superscriptaddress,showpacs]{revtex4-1}

\usepackage{graphicx}% Include figure files
\usepackage{epstopdf}
\usepackage[colorlinks=true,linkcolor=black, citecolor=blue, urlcolor=blue, 
    unicode=true]{hyperref}

\begin{document}

\title{Orphan spins and bound in-gap states in the $S={5\over2}$ antiferromagnet CaFe$_{2}$O$_{4}$}

\author{C. Stock}
\affiliation{School of Physics and Astronomy and Centre for Science at Extreme Conditions, University of Edinburgh, Edinburgh EH9 3FD, UK}
\author{E. E. Rodriguez}
\affiliation{Department of Chemistry and Biochemistry, University of Maryland, College Park, Maryland 20742, USA}
\author{N. Lee}
\affiliation{Rutgers Center for Emergent Materials and Department of Physics and Astronomy, Rutgers University, 136 Frelinghuysen Road, Piscataway, New Jersey 08854, USA}
\author{F. Demmel}
\affiliation{ISIS Facility, Rutherford Appleton Labs, Chilton, Didcot, OX11 0QX, UK}
\author{P. Fouquet}
\affiliation{Institute Laue-Langevin, 6 rue Jules Horowitz, Boite Postale 156, 38042 Grenoble Cedex 9, France}
\author{M. Laver}
\affiliation{Laboratory for Neutron Scattering, Paul Scherrer Institut, CH-5232 Villigen, Switzerland}
\author{Ch. Niedermayer}
\affiliation{Laboratory for Neutron Scattering, Paul Scherrer Institut, CH-5232 Villigen, Switzerland}
\author{Y. Su}
\author{K. Nemkovski}
\affiliation{J\"ulich Centre for Neuton Science JCNS, Forschungszentrum J\"ulich GmbH, Outstation at MLZ, Lichtenbergstra\ss e 1, D-85747 Garching, Germany}
\author{M. A. Green}
\affiliation{School of Physical Sciences, University of Kent, Canterbury, CT2 7NH, UK}
\author{J. A. Rodriguez-Rivera}
\affiliation{NIST Center for Neutron Research, National Institute of Standards and Technology, 100 Bureau Drive, Gaithersburg, Maryland, 20899, USA}
\affiliation{Department of Materials Science, University of Maryland, College Park, Maryland 20742, USA}
\author{J. W. Kim}
\affiliation{Rutgers Center for Emergent Materials and Department of Physics and Astronomy, Rutgers University, 136 Frelinghuysen Road, Piscataway, New Jersey 08854, USA}
\author{L. Zhang}
\affiliation{Laboratory for Pohang Emergent Materials and Max Plank POSTECH Center for Complex Phase Materials, Pohang University of Science and Technology, Pohang 790-784, Korea}
\author{S. -W. Cheong}
\affiliation{Rutgers Center for Emergent Materials and Department of Physics and Astronomy, Rutgers University, 136 Frelinghuysen Road, Piscataway, New Jersey 08854, USA}

\date{\today}

\begin{abstract}

CaFe$_{2}$O$_{4}$ is an anisotropic $S={5\over 2}$ antiferromagnet with two competing $A$ ($\uparrow \uparrow \downarrow \downarrow$) and $B$ ($\uparrow \downarrow \uparrow \downarrow$) magnetic order parameters separated by static antiphase boundaries at low temperatures.  Neutron diffraction and bulk susceptibility measurements, show that the spins near these boundaries are weakly correlated and a carry an uncompensated ferromagnetic moment that can be tuned with a magnetic field.  Spectroscopic measurements find these spins are bound with excitation energies less than the bulk magnetic spin-waves and resemble the spectra from isolated spin-clusters.  Localized bound orphaned spins separate the two competing magnetic order parameters in  CaFe$_{2}$O$_{4}$.

\end{abstract}


\maketitle

Coupling different order parameters often results in new states near the boundary separating them.~\cite{Shockley39:56,Mills68:171,Mills68:20}  This has been exploited in a variety of fields to engineer unusual properties including in the area of photonics.~\cite{Li03:90,Enoch02:89}  An example also occurs in the vortex state of superconductors where vortices host bound electronic states that differ from the bulk parent metal.~\cite{Overhauser89:62,Hess89:62,Hayashi98:80,Pan00:85}  Fermionic states that exist near boundaries can also be topologically protected~\cite{Hasan10:82} resulting in low-energy modes that are robust owing to a symmetry of the underlying Hamiltonian. Examples of such states occur near solitons in polyacetylene~\cite{Su80:77,Heeger88:60,Roth87:36,Rotherberg86:57}.  However, analogous boundaries and states in magnets, particularly antiferromagnets, have been difficult to identify owing to the absence of a net magnetization, fast dynamics, and the different statistics obeyed by bosonic magnons.\cite{Kimel04:429,Jungwirth16:11,Shiino16:117,Gomonay16:117,Kampfrath11:5,Bode06:5,Meikle56:102,Meikle57:105,Zhang13:87}   Here we investigate edge states in the classical and anisotropic antiferromagnetic CaFe$_{2}$O$_{4}$ near the boundary between two competing magnetic order parameters.

CaFe$_{2}$O$_{4}$ is a $S$=${5\over2}$ antiferromagnet with an orthorhombic space group (\#62 $Pnma$, $a$=9.230 \AA, $b$=3.017 \AA, $c$=10.689 \AA).~\cite{Decker57:10,Hill56:9,Allain66:9,Watanabe67:22,Das16:16}  The magnetic structure consists of two competing spin arrangements, termed the $A$ and $B$ phases (denoted as ($\uparrow \uparrow \downarrow \downarrow$) ($\uparrow \downarrow \uparrow \downarrow$) respectively), which are distinguished by their $c$-axis stacking of ferromagnetic $b$-axis stripes.~\cite{Obata13:121}  Neutron inelastic scattering has found that the magnetic exchange coupling in CaFe$_{2}$O$_{4}$ is predominately two dimensional with strong coupling along $a$ and $b$ compared to that along $c$.  Neutron diffraction has found that the two $A$ ($\uparrow \uparrow \downarrow \downarrow$) and $B$ ($\uparrow \downarrow \uparrow \downarrow$) magnetic phases both exist at low temperatures in single crystals and are separated by antiphase boundaries that are confining and result in a countable heirarchy of discrete magnetic excitations.~\cite{Stock16:117}  

\begin{figure*}[t]
\includegraphics[width=16cm] {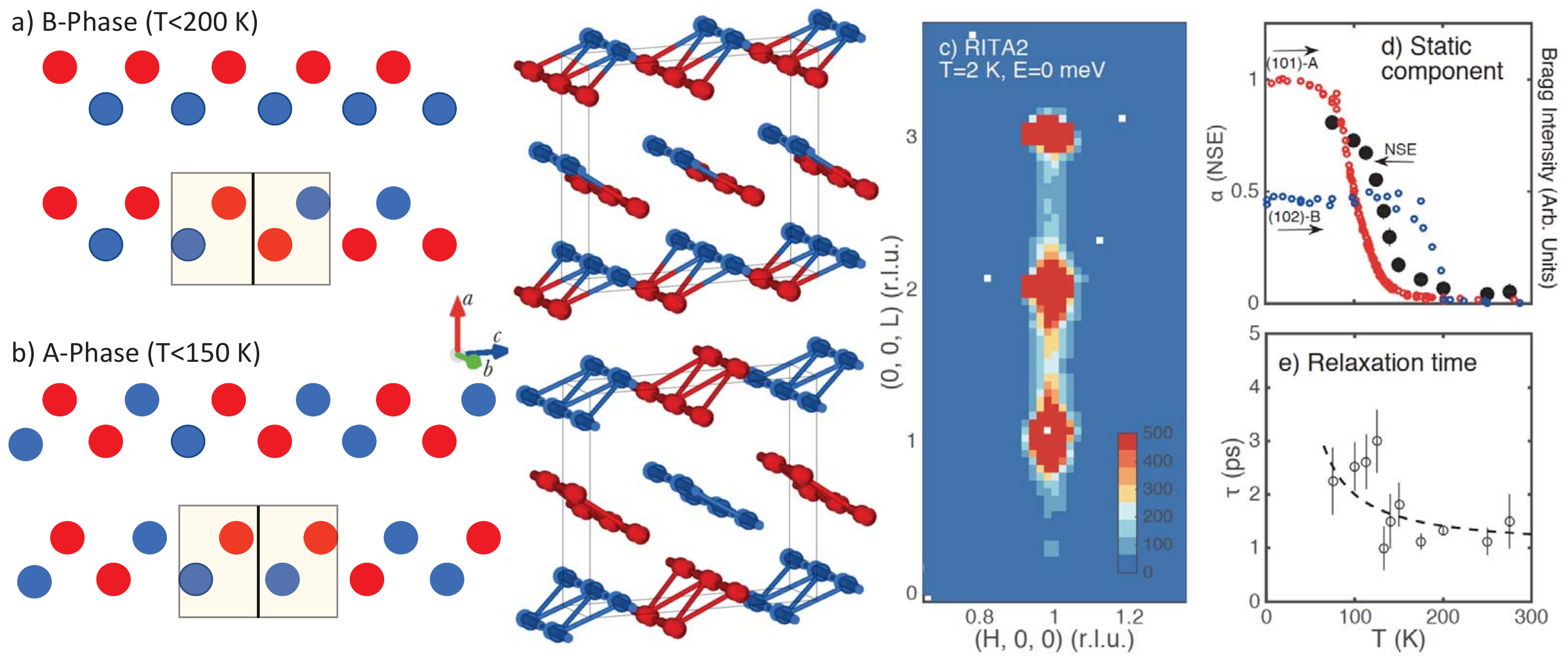}
\caption{\label{summary} $(a)$ Illustrates the magnetic $B$ phase ($\uparrow \downarrow \uparrow \downarrow$) of CaFe$_{2}$O$_{4}$ showing an antiphase boundary where locally (within the highlighted box) the magnetic structure is the $A$ phase ($\uparrow \uparrow \downarrow \downarrow$).  Note that this antiphase boundary carries a net ferromagnetic moment.  $(b)$ the same is illustrated for the low temperature $A$-phase where locally the magnetic structure is the $B$ phase.  $(c)$ illustrates the diffuse scattering cross section characterizing antiphase boundaries.  $(d-e)$ show results of a spin echo analysis plotting the fraction of static (on the $\sim$ GHz timescale) boundaries and the decay time.  The magnetic order parameters of the $A$ and $B$ phases extracted from neutron diffraction are also plotted.  The dashed line is discussed in the main text.}
\end{figure*}

The $A$ ($\uparrow \uparrow \downarrow \downarrow$) and $B$ ($\uparrow \downarrow \uparrow \downarrow$) magnetic structures are illustrated in Fig. \ref{summary} $(a)$ and $(b)$ with the magnetic moments aligned along the $b$ axis (antiparallel arrangements denoted as red and blue). Two possible antiphase boundaries along the $c$-axis are also illustrated.  In panel $(a)$, the boundary separates two high temperature $B$  ($\uparrow \downarrow \uparrow \downarrow$)  phase structures and locally has the magnetic structure of the low temperature  $A$ phase ($\uparrow \uparrow \downarrow \downarrow$) and also carries a net ferromagnetic moment.  A similar situation is presented in panel $(b)$ for the low temperature $A$ phase.  The momentum broadened rod of diffuse scattering characterizing these boundaries is reproduced in panel $(c)$.~\cite{Stock16:117} 

High resolution neutron spectroscopy (Fig. \ref{summary} $d$ which plots the static fraction $\alpha$ as a function of temperature) finds these boundaries are predominately static on the GHz timescale below $\sim$ 100 K.  The freezing of the boundaries occurs below the onset of $B$ phase ($\uparrow \downarrow \uparrow \downarrow$) order measured by the (102) magnetic Bragg peak and also higher then the onset of $A$ phase ($\uparrow \uparrow \downarrow \downarrow$) order probed through measurements of (101).  The relaxational timescale measured with spin echo is displayed in panel $(e)$ where the dashed line is a plot of $\tau = \exp(U/k_{B}T)$ with $U$ fixed at the bulk magnetic anisotropy gap of 5 meV measured with neutron spectroscopy.~\cite{Stock16:117}  The  data is consistent with antiphase boundaries relaxing with an energy fixed by the bulk spin anisotropy.

The presence of static boundaries separating $A$ and $B$ order parameters brings the possibility of magnetic states that have different properties from the bulk, termed orphan spins.~\cite{Moessner99:83,Schiffer97:56,Rehn17:118}  We apply neutron diffraction and inelastic scattering to identify and characterize these states.  Further experimental details are provided in the supplementary information.

\begin{figure}[t]
\includegraphics[angle=0,width=8.3cm] {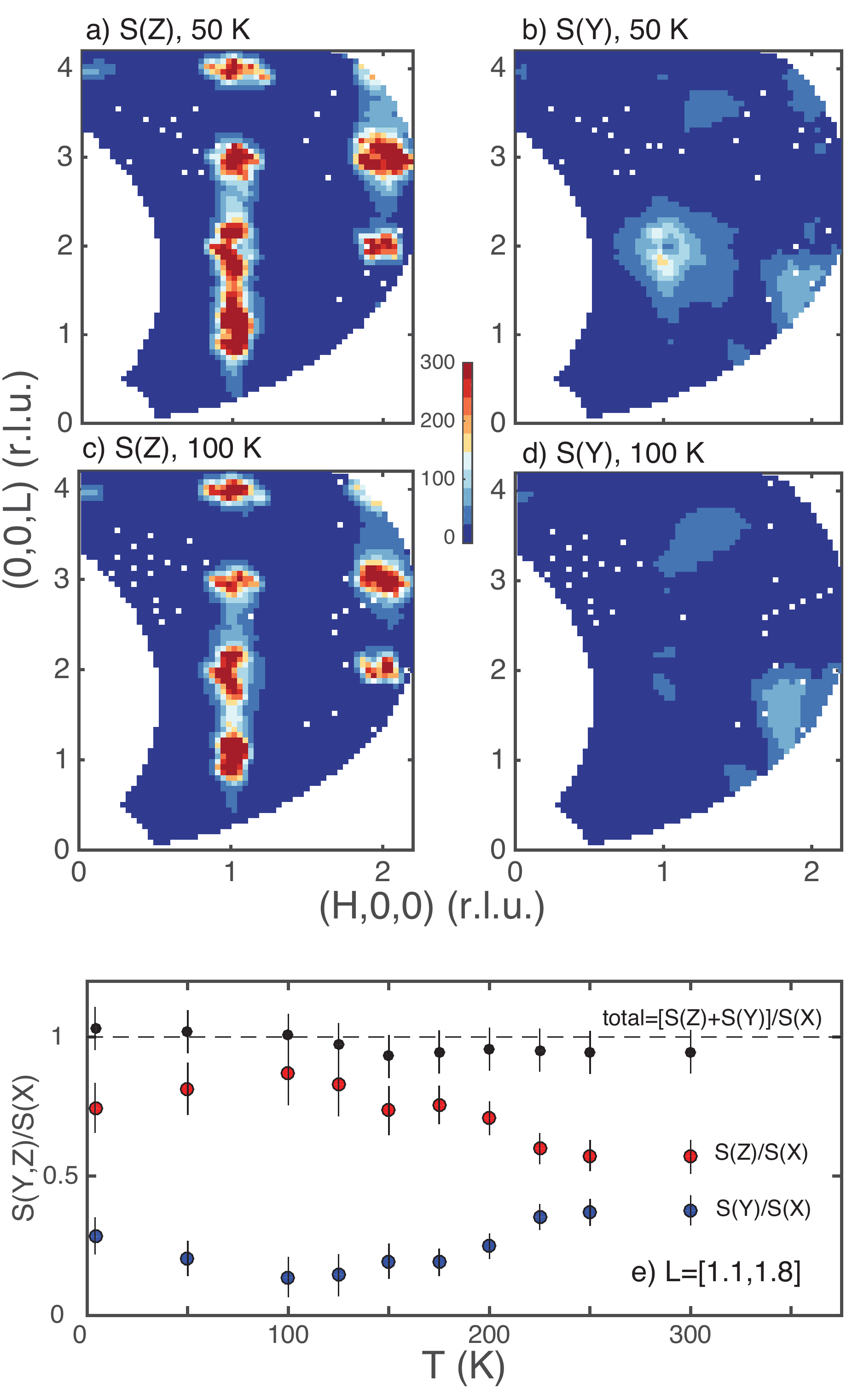}
\caption{\label{dns_summary}  Polarized (magnetic) diffuse scattering where  X is parallel to $\vec{Q}$, Y iperpendicular and within the (H0L) plane, and Z along $b$.  Panels $(a)$ and $(c)$ show the magnetic scattering originating from spins aligned along the crystallographic $b$ axis at 100 and 50 K.  $(b)$ and $(d)$ show the same but for the spins oriented perpendicular to $b$.  $(e)$ plots the fraction of intensity originating from spins aligned along Y and Z.  The total is shown to be in agreement of 1, required from sum rules for polarized neutron scattering.}
\end{figure}

We first investigate the static structure of the antiphase boundaries using the DNS polarized diffractometer applying an $XYZ$ polarization geometry.  Fig. \ref{dns_summary} illustrates the background corrected magnetic scattering originating from Fe$^{3+}$ moments pointing along $Y$ and $Z$ (with $Z$ vertical and parallel to the crystallographic $b$ axis and $Y$ in the horizontal (H0L) scattering plane and perpendicular to $\vec{Q}$).  Panels $(a-d)$ plot the magnetic intensity at 100 K and 50 K displaying two components - momentum resolution limited Bragg peaks at the integral (H,0,L) positions, corresponding to the long-range bulk structure, and a component which is broadened along the (1, 0, L) direction originating from short range spin correlations associated with the antiphase boundaries.  The intensity contours illustrate that while most of the low-temperature magnetic scattering originates from spins aligned parallel to the $b$-axis (Z direction), there is a measurable momentum broadened fraction of the intensity originating from moments perpendicular to this direction along $Y$.   Panel $(e)$ plots the temperature evolution of the two components divided by the total magnetic intensity from the $X$ direction showing a significant fraction of spins jam perpendicular to the crystallographic $b$-axis while the Fe$^{3+}$ moments reorient from $B$ ($\uparrow \downarrow \uparrow \downarrow$) phase to $A$ ($\uparrow \uparrow \downarrow \downarrow$)  phase order on cooling.  The polarized results illustrate that there is a gradual change in the spin direction across the domain wall reminiscent of a ``Bloch" wall instead of a fully discontinuous 180$^{\circ}$ ``Neel" type boundary. 

\begin{figure}[t]
\includegraphics[width=8.5cm] {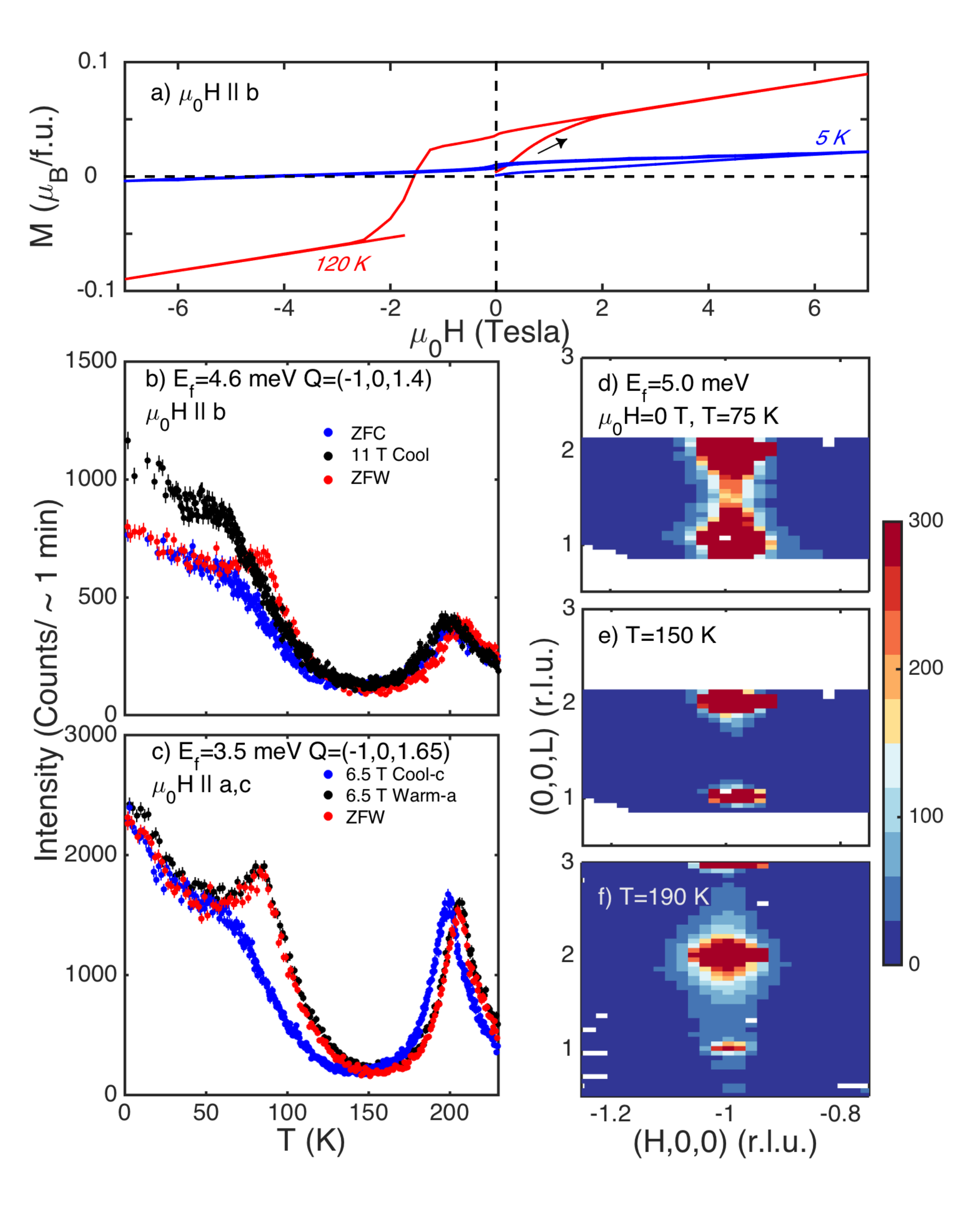}
\caption{\label{diffuse_summary}  $(a)$ Magnetization loops at 120 K and 5 K illustrating a remanent magnetization in CaFe$_{2}$O$_{4}$.  $(b)$ plots the temperature dependence at $\vec{Q}$=(-1, 0, 1.4).  Zero field cooled (ZFC), zero field warmed (ZFW), and 11 T field cooled data were taken with the field $||$ $b$ (vertical) axis.  $(c)$ illustrates the same cooling sequences at  $\vec{Q}$=(-1, 0, 1.65) with the field aligned perpendicular to the $b$-axis and in the (H0L) scattering plane.  $(d-f)$ plots the diffuse scattering cross section in zero field at 75 K, 150 K, and 190 K.  Further details on the experimental configuration and zero field susceptibility are given in the supplementary information.}
\end{figure}

We now investigate whether these boundaries are tunable with an applied magnetic field.~\cite{Papanicolaou98:10}  Magnetization loops at 120 K and 5 K in Fig. \ref{diffuse_summary} $(a)$ find an uncompensated remanent moment when the field is applied along $b$.  Panels $(b,c)$ illustrated the temperature and magnetic field dependence of the elastic diffuse scattering (RITA2 with unpolarized neutrons) at $\vec{Q}$=(-1, 0, 1.4) and (-1, 0, 1.65) under different applied field conditions and  representative (H0L) maps are displayed in panels $(d-f)$.  The peak in intensity at $\sim$ 200 K (panels $b,c$) is associated with critical scattering of the high temperature $B$ phase ordering (panels $d-f$).  A minimum in the temperature dependent intensity (panels $b,c$) is seen at $\sim$ 150 K  before rod like scattering along L characteristic of static antiphase boundaries forms (panel $(d)$ at 75 K).  Panels $(b,c)$ show that the intensity is hysteretic in temperature with a peak forming at $\sim$ 100 K on warming analogous to localized structures in disordered materials (for example ferroelectric K$_{1-x}$Li$_{x}$TaO$_{3}$~\cite{Stock14:90}).

Fig. \ref{diffuse_summary} $(b)$ and $(c)$ also display the temperature dependence of this diffuse scattering cross section in the case of differing field conditions.  When cooling takes place in a 11 T field parallel to the $b$ axis (panel $b$), the diffuse scattering is enhanced in comparison to the zero field cooled (ZFC) temperature sweep.   No field dependence in this enhancement was observed for $\mu_{0}H$ greater than 1 T and the effect was observed to freeze in for cooling below $\sim$ 150 K.  Panel $(c)$ illustrates that this enhancement is largely reduced when the field is perpendicular to the $b$ axis as shown using a horizontal magnetic field of 6.5 T.   The comparatively small changes with the field perpendicular to the $b$ axis is consistent with the relatively small number of spins jammed perpendicular to $b$ discussed above in the context of Fig. \ref{dns_summary}.  Due to kinematic constraints associated with the horizontal magnet, an E$_{f}$=3.5 meV was used providing different intensity ratios for the diffuse scattering measured at $\sim$ 200 K in comparison to base temperature owing to differing energy resolutions and spectrometer configurations.   Therefore, cooling with the field aligned along the direction of dominant bulk staggered magnetization (crystallographic $b$ axis) results in an enhancement of diffuse scattering indicative of a larger density of antiphase boundaries.  Orienting the field perpendicular does not result in any such enhancement.  

The response of the diffuse scattering to an applied magnetic field that tracks the dominant orientation indicates that these boundaries have a $b$-axis uncompensated, ferromagnetic, moment.  While this conclusion is drawn from the finite-$Q$ response, magnetization (panel $a$, $Q=0$ probe) corroborates the presence of a localized ferromagnetic moment and further data presented in the supplementary information show the momentum dependence is indeed peaked at $Q$=0.  One such real-space scenario for this to occur is illustrated in Fig. \ref{summary} $(a)$ which schematically plots an antiphase boundary in the high temperature $B$ phase.  Locally, the orphaned spins in the boundary have the structure of the low temperature $A$ phase and also carry a net ferromagnetic moment which originates in the field dependence presented in Fig. \ref{diffuse_summary}.   This local ferromagnetism occurs even though the magnetic structure is globally antiferromagnetic. 

\begin{figure}[t]
\includegraphics[width=8.8cm] {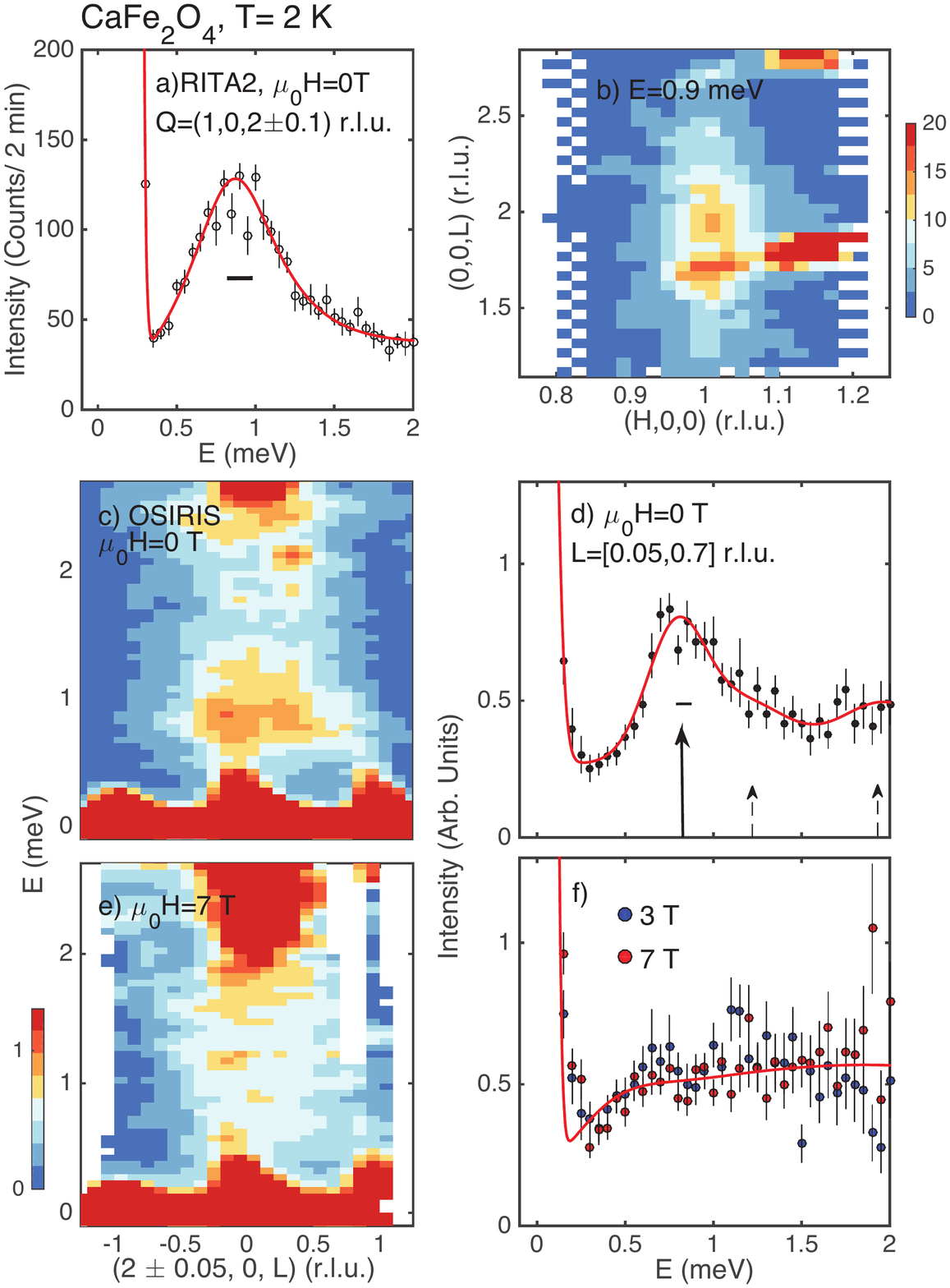}
\caption{\label{magfield}  $(a)$ displays a constant-Q scan showing the presence of an in-gap mode at low temperatures fit to a harmonic oscillator lineshape.  $(b)$ illustrates a constant energy slice showing that the intensity is elongated along L.  $(c-f)$ displays high resolution scans taken on OSIRIS showing the gapped excitation and the response to 3 and 7 T applied along the crystallographic $b$-axis.  The solid line in $(d)$ is a fit to underdamped harmonic oscillators with positions fixed to be the calculated dimer (solid arrow) and trimer (dashed arrows) positions.  The fit in $(f)$ is to a single damped harmonic oscillator.}
\end{figure}

We now apply spectroscopy to study the energy spectra associated with these antiphase boundaries.  Magnetic boundary, or edge, states have been predicted in low dimensional magnets~\cite{Ng94:50} and superconductors~\cite{Kohno99:68} and experimentally observed in insulating and disordered quantum magnets~\cite{Xu07:317,Kenzelmann03:90,Els98:43,Schmidiger16:116,Stock09:79}.  Motivated by the possibility of novel states near these boundaries, we apply neutron spectroscopy in Fig. \ref{magfield} by searching for bound magnetic excitations within the anisotropy induced gap of $\sim$ 5 meV.  Fig. \ref{magfield} $(a)$ illustrates a constant momentum scan (RITA2) showing a peak at 0.9 meV.  The peak is significantly broader than resolution (solid horizontal line of 0.25 meV) with a full width in energy of $2\Gamma$=0.72 $\pm$ 0.15 meV and approximately an order of magnitude weaker in intensity than the bulk dispersive spins waves.    Panel $(b)$ plots a constant energy slice indicating strong correlations along the $a$ axis and weak correlations along $c$ mimicking the elastic magnetic diffuse scattering cross section (Fig. \ref{summary} $c$).  Figure \ref{magfield} panel $(c)$ shows an energy slice using high resolution neutron spectroscopy from the OSIRIS backscattering spectrometer.  The mode at 0.9 meV, while broader than resolution, displays no momentum dispersion and hence no on-site molecular field, indicative of isolated or orphan spins states.

The energy scale of 0.9 meV can be reconciled if we consider a simple edge state consisting of isolated clusters.  Such clusters consist of $S={5\over 2}$ Fe$^{3+}$ spins coupled with an exchange constant along the crystallographic $c$ axis with an interaction Hamiltonian of $H=J_{c}\sum_{ij} \vec{S}_{i}\cdot\vec{S}_{j}$ (where $i,j$ is summed over the cluster).~\cite{Buyers84:30,Furrer13:85}  The simplest state would consist of an isolated dimer with a singlet $j_{eff}$=0 ground state and higher energy levels of $j_{eff}$=1,2,3,4,5.  The energy scale to excite such a dimer from the ground state to an excited state is $=J_{c}$ which has been estimated to be 0.94 $\pm$0.19 meV based on high energy spectroscopy of the bulk magnetic dispersion discussed previously.~\cite{Stock16:117}  This is in agreement with the peak position in Fig. \ref{magfield} $(a)$.   However, Fig. \ref{magfield} $(c)$ also displays a continuum of excitations that extend from E=0.9 meV to higher energies which can be understood in terms of larger clusters such as trimers which would display discrete excitations at further energies.  The energy spectrum for the above Hamiltonian based on a trimer would display lowest excitation energies of 1.5, 2.5, and 3.5$J_{c}$.~\cite{Svensson78:49}   The solid line in panel $(d)$ is a fit to the OSIRIS data to a series of lifetime shortened excitations fixed at the dimer excitation level and the two lowest energy trimer levels with the intensity reflecting the probability of such states.   From this fit to dimer and trimers, an estimate of $J_{c}$=0.78 $\pm$ 0.17 meV which is in agreement with the value obtained from fitting the dispersive band excitations.

Fig. \ref{magfield} $(e)$ and $(f)$ illustrate the response of these cluster states to an applied magnetic field showing that applied fields of 3 and 7 T along the crystallographic $b$ axis are sufficient to smear the lowest energy state in energy.  These results are consistent with Zeeman splitting of lifetime shortened multiplets originating from cluster excitations.  The fit in panel $(f)$ is to a single energy broadened relaxational mode.  The results of this analysis shows that the exchange constant derived from higher energy bulk spin wave measurements and the localized excitations from the ``in-gap" states can be consistently understood by the presence of clusters of spins located near the antiphase boundaries.  The energy scale of these cluster states is low enough to be tuned with a field.

The magnetic bound states display weak dynamic correlation lengths along $c$, while much longer length scales along $a$, therefore mimicking the planar antiphase boundaries found in diffraction and differing from the spin-waves onset at much higher energies.   The lack of a measurable on-site molecular field evidenced from the momentum dependence indicates that these orphaned spins are decoupled from the $A$ and $B$ magnetic order parameters.  These orphaned states exist at the boundary between the two order parameters allowing them to coexist in CaFe$_{2}$O$_{4}$ at low temperatures.  Such states have been proposed as a means of stabilizing spin liquid states in honeycomb lattices~\cite{Flint13:11} and may exist in triangular magnets with much smaller exchange interactions resulting in strong low-energy fluctuations.~\cite{Lhotel11:107,Stock10:105,Nambu15:115}.  Orphaned spins maybe a means of decoupling differing magnetic orders when a number of different order parameters exist with similar energy scales.

In summary, we have shown the presence of ferromagnetic edge states in CaFe$_{2}$O$_{4}$ originating from antiphase boundaries separating competing magnetic order parameters.  Spectroscopic evidence points to these edge states consisting of clusters of orphaned spins.

\begin{acknowledgments}
This work was supported by the EPSRC, the Carnegie Trust for the Universities of Scotland, the Royal Society of London, and Royal Society of Edinburgh, the STFC, EU-NMI3, NSF (No. DMR-1508249), and the Swiss spallation neutron source (SINQ) (Paul Scherrer Institute, Villigen, Switzerland).  The work at Rutgers was supported by the DOE under Grant No. DE-FG02-07ER46382.  The work at Postech was supported by the Max Planck POSTECH/KOREA Research Initiative Program [Grant No. 2011-0031558] through NRF of Korea funded by MSIP.  We are grateful to C. Mudry for helpful discussions.
\end{acknowledgments}

%\bibliography{CFO_bib}


\end{document}

\title{Supplementary Information: ``Orphan spins and bound in-gap states in the $S={5\over2}$ antiferromagnet CaFe$_{2}$O$_{4}$"}

\author{C. Stock}
\affiliation{School of Physics and Astronomy and Centre for Science at Extreme Conditions, University of Edinburgh, Edinburgh EH9 3FD, UK}

\author{E. E. Rodriguez}
\affiliation{Department of Chemistry and Biochemistry, University of Maryland, College Park, Maryland 20742, USA}

\author{N. Lee}
\affiliation{Rutgers Center for Emergent Materials and Department of Physics and Astronomy, Rutgers University, 136 Frelinghuysen Road, Piscataway, New Jersey 08854, USA}

\author{M. A. Green}
\affiliation{School of Physical Sciences, University of Kent, Canterbury, CT2 7NH, UK}

\author{F. Demmel}
\affiliation{ISIS Facility, Rutherford Appleton Labs, Chilton, Didcot, OX11 0QX, UK}

\author{P. Fouquet}
\affiliation{Institute Laue-Langevin, 6 rue Jules Horowitz, Boite Postale 156, 38042 Grenoble Cedex 9, France}

\author{M. Laver}
\affiliation{Laboratory for Neutron Scattering, Paul Scherrer Institut, CH-5232 Villigen, Switzerland}

\author{Ch. Niedermayer}
\affiliation{Laboratory for Neutron Scattering, Paul Scherrer Institut, CH-5232 Villigen, Switzerland}

\author{Y. Su}
\author{K. Nemkovski}
\affiliation{J\"ulich Centre for Neuton Science JCNS, Forschungszentrum J\"ulich GmbH, Outstation at MLZ, Lichtenbergstra\ss e 1, D-85747 Garching, Germany}

\author{J. A. Rodriguez-Rivera}
\affiliation{NIST Center for Neutron Research, National Institute of Standards and Technology, 100 Bureau Drive, Gaithersburg, Maryland, 20899, USA}
\affiliation{Department of Materials Science, University of Maryland, College Park, Maryland 20742, USA}

\author{S. -W. Cheong}
\affiliation{Rutgers Center for Emergent Materials and Department of Physics and Astronomy, Rutgers University, 136 Frelinghuysen Road, Piscataway, New Jersey 08854, USA}

\date{\today}

\begin{abstract}

Supplementary information is provided giving experimental details, details regarding the magnetic susceptibility, horizontal field measurements, further coarser resolution data, and calculations for dimer and trimer states.  

\end{abstract}


\maketitle

\renewcommand{\thefigure}{S1}
\begin{figure}[t]
\includegraphics[width=8.5cm] {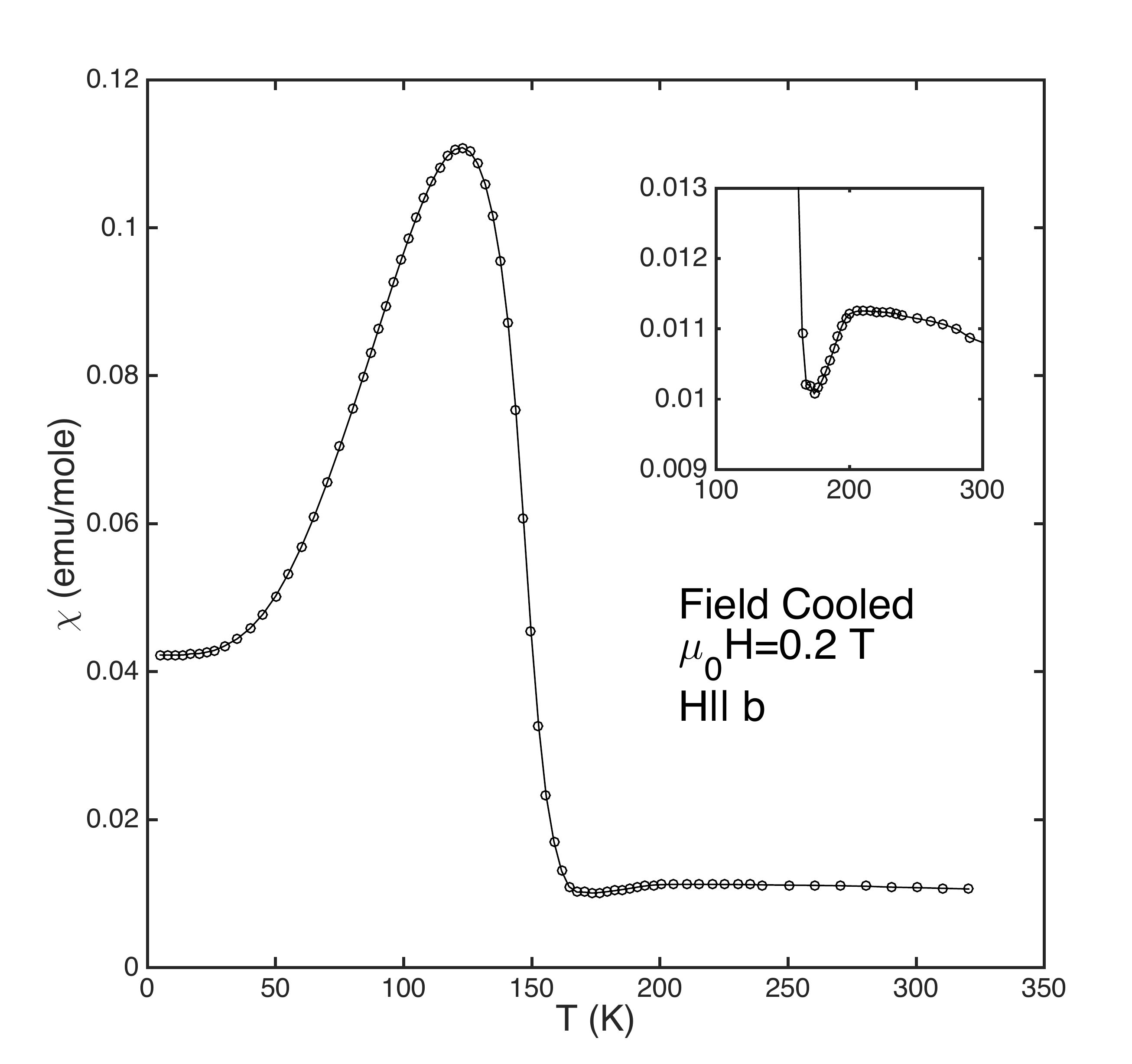}
\caption{\label{chi}  Susceptibility measured with the field aligned along the $b$ axis and field cooled in a 0.2 T field.  The inset shows an enhanced plot near 200 K to illustrate the anomaly at this temperature.}
\end{figure}

\renewcommand{\thefigure}{S2}
\begin{figure}[t]
\includegraphics[width=8.5cm] {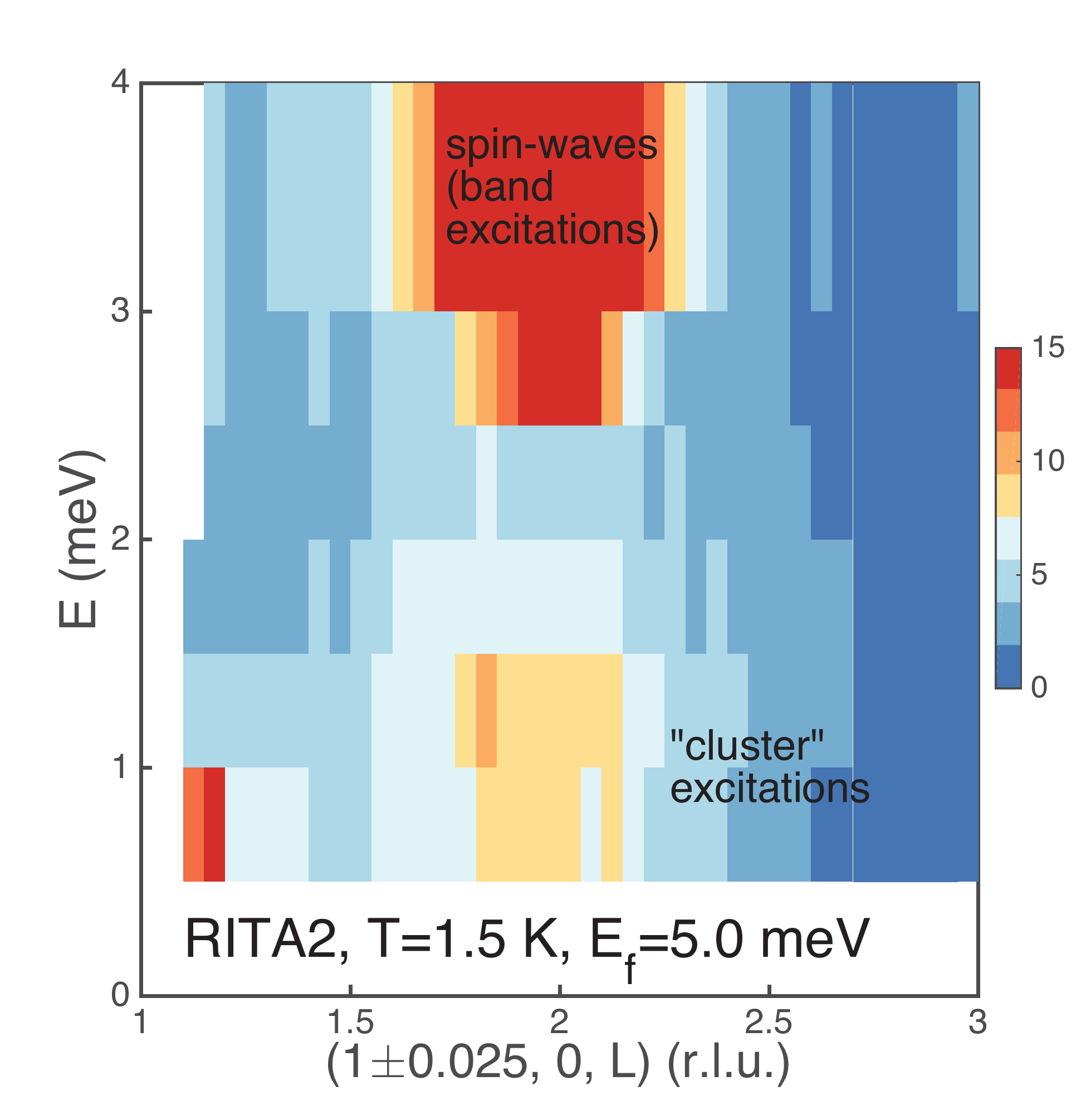}
\caption{\label{coarse}  Constant-Q slice taken on the RITA2 spectrometer with E$_{f}$=5.0 meV giving coarser energy resolution than used in the main text, but providing a broader dynamic range.  The gapped dispersive ``band" excitations are annotated at higher energies, while the observed cluster excitation is observable at low-energies.  This mode was studied with finer energy resolution in the main text.}
\end{figure}

\renewcommand{\thefigure}{S3}
 \begin{figure}[t]
\includegraphics[width=9cm] {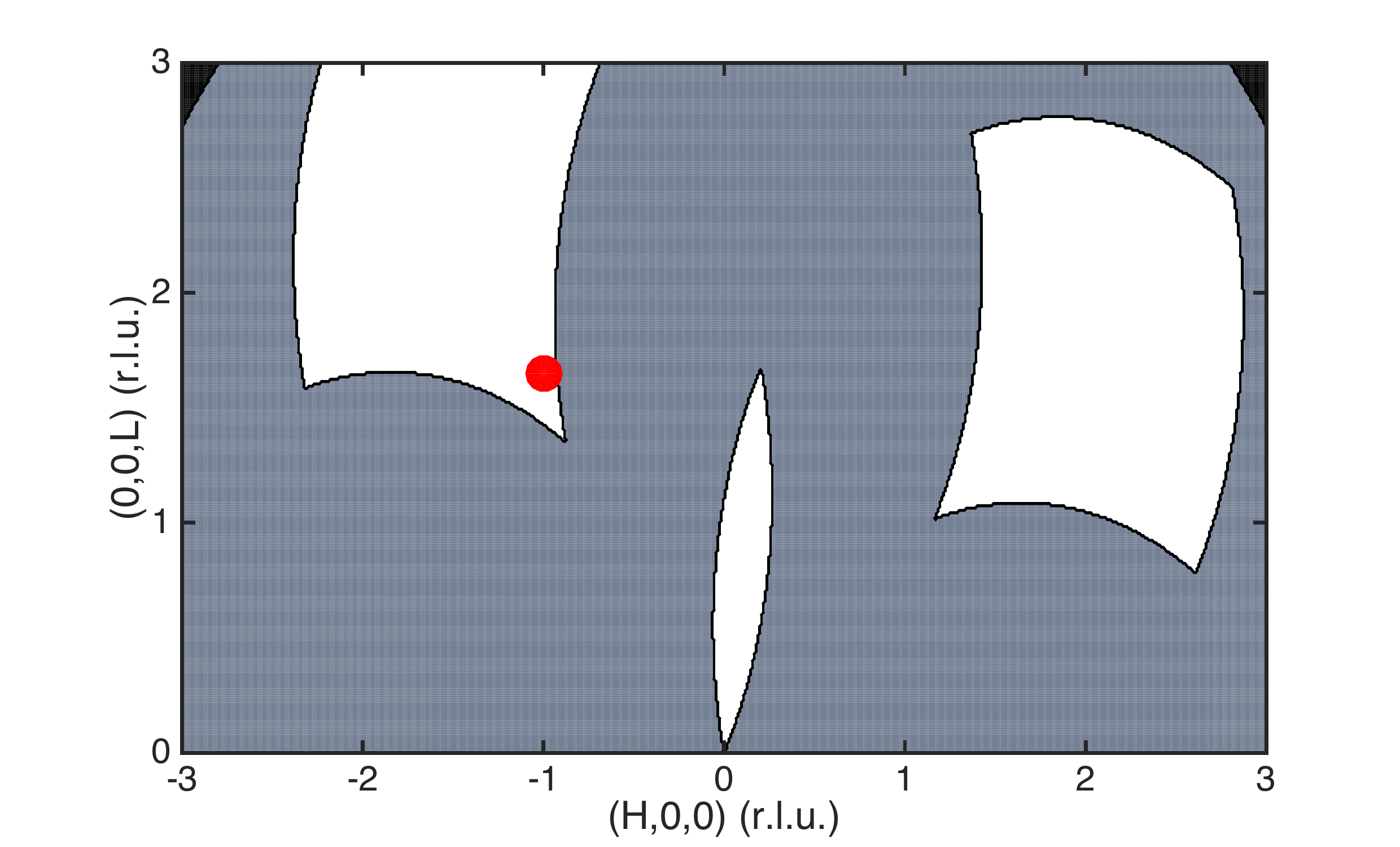}
\caption{\label{phase_space}  The allowed momentum positions with the horizontal magnet with 4 $\times$ 45$^{\circ}$ windows.  The white areas are the regions in reciprocal space where the beam can be enter and leave the magnet.  The grey regions are those kinematically allowed, but result in a beam collision with the magnet.  The black regions are momentum positions not kinematically allowed.  The red dot was the position chosen to perform the field dependent scans.}
\end{figure}

\textit{Susceptibility and characterisation:}  Sample quality was confirmed using susceptibility, x-ray powder diffraction, and further neutron diffraction.  Magnetic susceptibility data is shown in Fig. \ref{chi} and illustrates the two magnetic transitions found using neutron diffraction (Fig. 1 of the main text) when the field is aligned along the $b$ axis of the single crystal.  The oxygen content was confirmed through x-ray and neutron powder diffraction studies of pieces from the single crystal and also on the starting materials for the single crystal. 

\textit{Experimental Details:}  The work presented in the main text was based upon measurements performed on the DNS diffractometer (FRM2, Germany), RITA-2 (PSI, Switzerland), SPINS and BT9 (NIST), and OSIRIS (ISIS, UK).  The sample used in all studies was made using the floating zone technique.

\textit{SPINS and  BT9 (NIST):} Single crystal diffraction data measuring the magnetic structure as a function of temperature were performed on the SPINS and BT9 triple-axis spectrometers (NIST, USA).  Vertically focussed PG002 graphite was used to produce a monochromatic beam incident on the sample and a flat PG002 graphite analyzer was used on the scattered side to select a final energy and reduce background.  For SPINS the final energy was fixed at E$_{f}$=5.0 meV and BT9 E$_{f}$=13.7 meV.  Cold beryllium filters were used to remove higher order contamination on SPINS.  A graphite filter was used on BT9 on both the incident and scattered sides.  For experiments done on both instruments, the collimation sequence was set to $open-80-S-80-open$.

\textit{RITA-2 (PSI):} Complementing our data from SPINS and BT9, we have also presented neutron elastic and inelastic data from RITA-2.  For these measurements, we have used the position sensitive detector which is well suited for tracking diffuse scattering and studying dispersive excitations.   Similar to SPINS, a vertically focussed PG002 monochromator was used along with a flat PG002 analyzer.  The final energy was fixed to E$_{f}$=5.0 meV, 4.6, and 3.5 meV as labelled in the figures in the main text.  The excitations were studied by varying E$_{i}$ therefore defining the energy transfer as $E=E_{i}-E_{f}$.  Below, a coarser resolution setup is used to illustrate both the cluster and band excitations.  The background from higher order scattering was reduced using a cold Beryllium filter on the scattered side.  This configuration gave an energy resolution (full-width) of $2\delta E$=0.21 meV at the elastic position.  Calibrations for the absolute moments of the $A$ and $B$ phases were done using RITA-2 by comparing the (1,0,1) and (1,0,2) Bragg peaks against 12 nuclear Bragg peaks where magnetic scattering was absent.  Crystallographic data for the nuclear structure factors was confirmed with powder experiments on the BT1 powder diffractometer at NIST.

\textit{DNS (FRM2)}: To confirm the magnetic structure and its relation to the magnetic structure, we used the DNS diffractometer at FRM2 (Germany).  Measurements were made with E$_{i}$=4.6 meV in two-axis mode (energy integrating) therefore providing an approximate measure $S(\vec{Q})$.  The incident beam was selected using a horizontally and vertically focused PG002 monochromator.  The beam was then polarized using m=3 Scharpf supermirror polarizers.  The polarization at the sample was fixed through the use of an $XYZ$ coil with the $X$ direction chosen to be parallel to the average $\vec{Q}$ at the sample and the $Z$ vertical aligned along the crystallographic $b$-axis.  The $Y$ direction was defined as being perpendicular to both $X$ and $Z$, and within the (H0L) scattering plane.  With the use of flipping coils in the incident and scattered beams, the two spin-flip and non-spin-flip cross sections could be measured with the neutron polarization along the three orthogonal Cartesian coordinates.  The flipping ratio was 20 $\pm$ 1 and was not found to deviate from this value regardless of the direction of neutron polarization.  All spin-flip data have been corrected for the feed-through from the non-spin-flip channel.  The scattered beam was measured with 24 detectors equally spaced $5^{\circ}$ apart covering a total angular range in scattering angle equal to 120$^{\circ}$.   

\renewcommand{\thefigure}{S4}
 \begin{figure}[t]
\includegraphics[width=7cm] {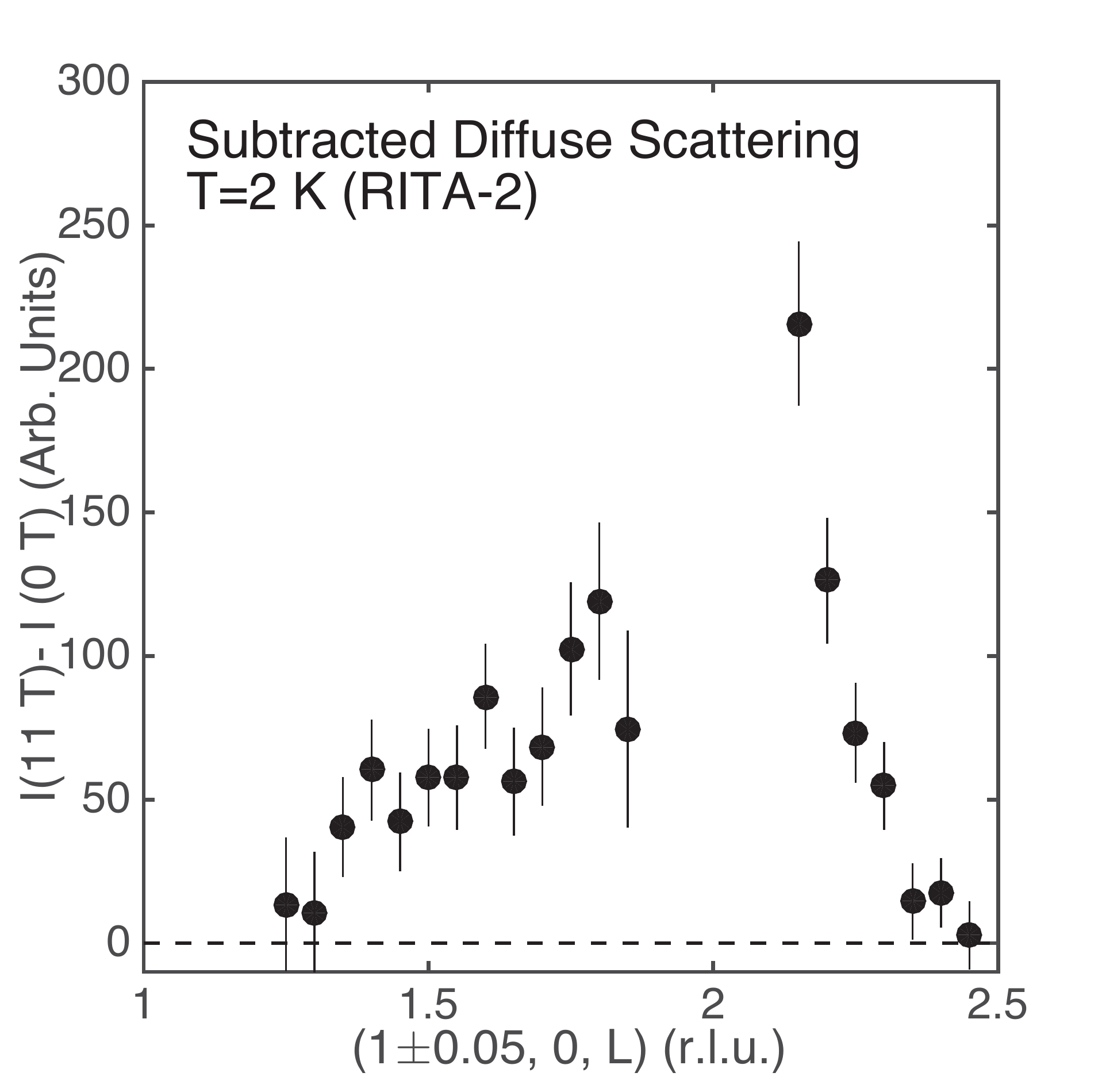}
\caption{\label{subtract_diffuse} The difference between 11 T and 0 T scans along the (1, 0, L) line.  The enhanced diffuse scattering can be seen to maximum at (1, 0, 2) and therefore associated with enhanced B phase ordering.}
\end{figure}

\renewcommand{\thefigure}{S5}
 \begin{figure}[t]
\includegraphics[width=6cm] {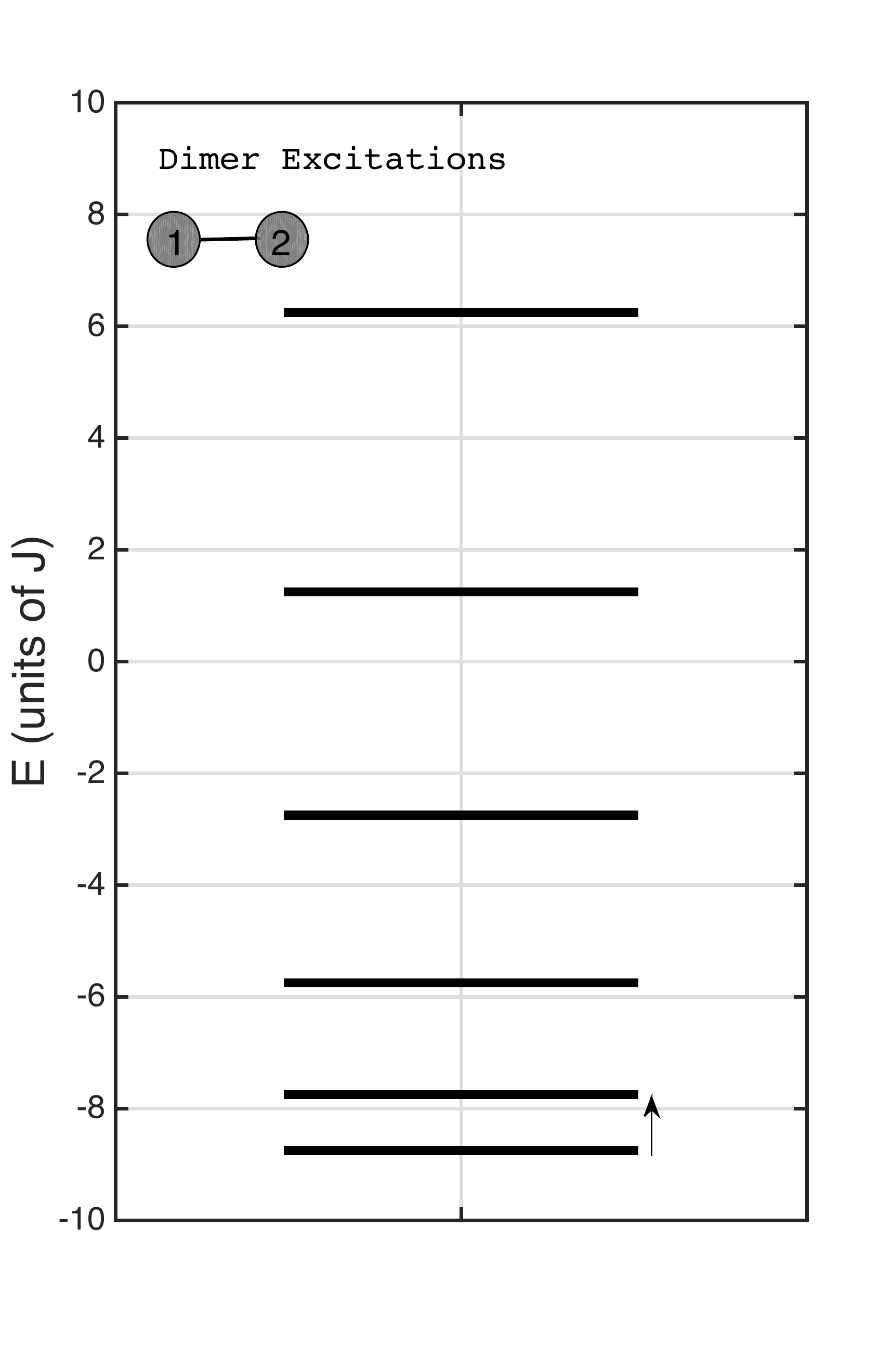}
\caption{\label{dimer}  Dimer energy levels in units of the exchange constant $J$ for $S={5\over 2}$ spins. Each level can be labelled with an effective total angular momentum characterizing degeneracy with $j_{eff}$=0 ... 5. }
\end{figure}

\renewcommand{\thefigure}{S6}
 \begin{figure}[t]
\includegraphics[width=6cm] {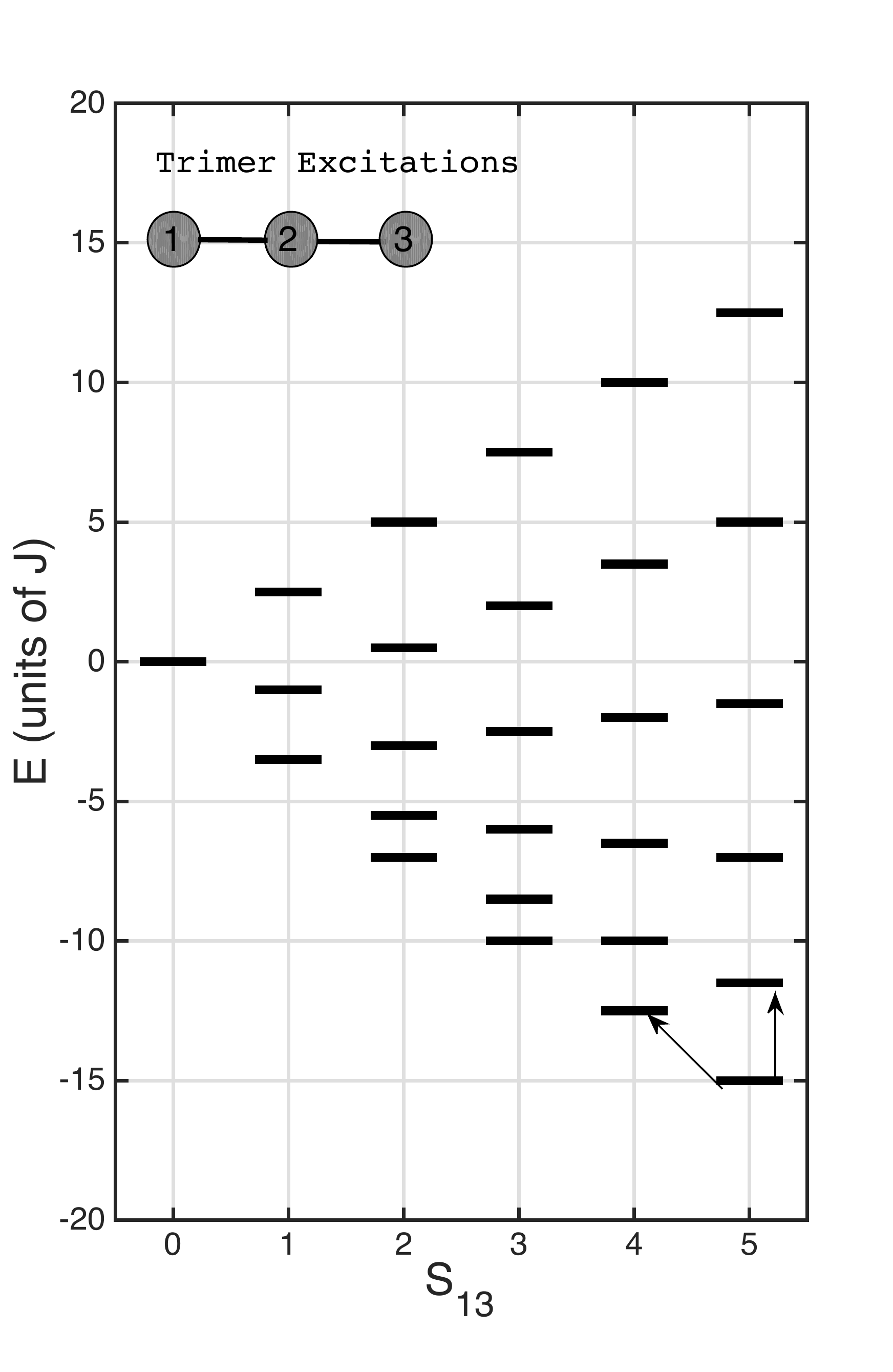}
\caption{\label{trimer}  Trimer energy levels in units of the exchange constant $J$.  The x-axis labels  $\vec{S}_{13}=\vec{S}_{1}+\vec{S}_{3}$ which extends from 0 to 5 for a trimer made up of $S={5\over 2}$ spins.}
\end{figure}

\textit{Polarized Neutron Analysis:}  During the DNS experiment, 6 different cross sections were measure with both spin-flip (SF) and non spin-flip cross sections measured for each of the $XYZ$ polarization directions.   The magnetic and background corrected magnetic cross section along the $XYZ$ directions, denoted as $F(X,Y,Z)$ in the main text, were obtained as listed below.

\begin{eqnarray}
F(X)=I_{SF,X}-I_{NSF,X} \nonumber \\
F(Y)=\left(\left(I_{SF,X}-I_{SF,Y}\right) +\left(I_{NSF,Y}-I_{NSF,X}\right)\right)/2 \nonumber \\
F(Z)=\left(\left(I_{SF,X}-I_{SF,Z}\right) +\left(I_{NSF,Z}-I_{NSF,X}\right)\right)/2 \nonumber
\nonumber
\end{eqnarray}

\textit{OSIRIS (ISIS):} High momentum and energy resolution data was obtained using the OSIRIS backscattering spectrometer located at ISIS.  A whitebeam of neutrons was incident on the sample and the final energy was fixed at E$_{f}$=1.84 meV using cooled graphite analyzers.  A cooled Beryllium filter was used on the scattered side to reduce background.  The default time configuration is set for a dynamic range of $\pm$ 0.5 meV, however by delaying the time channels, the dynamic range was extended to 3 meV.   An elastic energy resolution (full-width) of $2\delta E$=0.025 meV was obtained for these experiments. 
  
 \textit{IN11 (ILL):} Spin-echo measurements were done on the IN11 spectrometer (ILL, France) probing the low-energy $\sim$ GHz spin response.  Neutron spin-echo spectroscopy differs from other neutron methods in that it measures the real part of the normalized intermediate scattering.   This is achieved by encoding the neutron's speed into the Larmor precession of its nuclear magnetic moment in a well controlled, externally applied magnetic field.  $I(Q,t)$ is the spatial Fourier transform of the Van Hove self correlation function $G(r,t)$ which, essentially, gives the probability of finding a particle after time $t$ at a radius $r$ around the original position.   To obtain the dynamic range in time required, two wavelengths were used - 5.5 \AA\ and 4.2 \AA.  
 
\textit{Band and cluster excitations}:  In the main text, triple-axis data is presented illustrating the cluster excitations with higher resolution data presented from time of flight backscattering measurements.  To put these energy scales into context, a coarser resolution scan is presented in Fig. \ref{coarse}.  This constant-Q slice was obtained by tuning the spectrometer to have a E$_{f}$=5.0 meV while higher resolution data presented in the main text (Fig. 4) were taken with E$_{f}$=3.5 meV and with the high resolution backscattering OSIRIS.  This slice shows the bulk spin-waves at higher energy transfers and the low temperature cluster excitation at lower energies.   The spin-wave excitations characteristic of the bulk were presented in our previous study on confined spin excitations in CaFe$_{2}$O$_{4}$ and extend up $\sim$ 35 meV with strong dispersion measured in the $a-b$ plane and only weak dispersion along $c$.
 
\textit{Horizontal Magnetic fields and Triple-axis Spectrometers:}  To prove that the antiphase boundaries carry a ferromagnetic moment, it was necessary to align the field along the crystallographic $b$ axis and also perpendicular to it.  Measurements with the field aligned along $b$ observed a measurable enhancement with field, while those perpendicular (along the crystallographic $a$ and $c$ axes) did not observe an effect.  The second set of measurements with the field aligned along the crystallographic $a$ and $c$ axes required the use of a horizontal magnet given that the scattering plane was constrained to the (H0L) plane to study the diffuse scattering.  The choice of momentum positions for the temperature dependent scans presented in Fig. 3 in the main text were motivated based on several factors.  We chose the (-1, 0, 1.4) (for the $\mu_{0}H$ parallel to  $b$ study) position for our study to lie on the rod of diffuse scattering and also to be away from the half-integer positions to ensure there was no contamination from $\lambda/2$.  At the same time, we chose this position over (-1, 0, 1.6) to be further away in momentum from the large (-1, 0, 2) position.  To further prevent contamination from higher order wavelengths, in this configuration, we used an additional graphite filter in the incident beam.

The choice of momentum position for horizontal magnetic field measurements was more complicated.  The horizontal magnet consisted of 4 $\times$ 45$^{\circ}$ windows for the neutron beam to be incident and scattered from the sample.  The geometry is highly constraining in terms of reciprocal space and to reach a momentum position where strong diffuse scattering was present, the field had to be further tilted off-axis by 8$^{\circ}$ within the horizontal (H0L) plane.  The allowed phase space where the incident and scattered beam to not collide with the magnet is shown in Fig. \ref{phase_space}.  The red point is the position scanned and presented in the main text.  The final energy of E$_{f}$=3.5 meV was chosen to allow the diffuse scattering rods to enter into the allowed phase space constrained by the magnet and illustrated in Fig. \ref{phase_space}. Experimentally, no difference was observed with the field aligned either the $a$ or $c$ axes for field of up to 6.5 T.  This combination of factors based on kinematics fixed us to use the (-1, 0, 1.65) position for the horizontal field measurements.  As illustrated in Fig. \ref{subtract_diffuse}, the choice of momentum position did not change the the effect reported. 

\textit{Momentum dependence of enhanced field induced diffuse scattering:}  In the main text, we reported a field induced enhancement of diffuse scattering by performing temperature and field scans at the (-1, 0, 1.4) position.  Figure \ref{subtract_diffuse} shows the $L$ dependence of the diffuse scattering showing that it is peaked near the (1, 0, 2) position.  This indicates that this enhanced scattering is associated with local $B$ phase ($\uparrow \downarrow \uparrow \downarrow$) ordering.  An example of such an antiphase boundary is shown in Fig. 1 of the main text, antiphase boundaries associated with the B-phase carry a ferromagnetic moment and hence tunable with a magnetic field. 

\textit{Dimer and Trimer excitations:}   The energy scale of the in-gap excitation presented in the main text, matched that, within error, of the excitation energy expected for a dimer excitation.  The higher energy continuum could be further understood by weaker trimer excitations.  Fig. \ref{dimer} and \ref{trimer} illustrate the energy schematic for both types of excitations.  The dimer excitation levels can be labelled with an effective angular momentum of $j_{eff}$=0...5 which describes the degeneracy of each level.   The lowest level transition occurs with an excitation energy of $J$ and corresponds to a transition from $j_{eff}$=0 to $j_{eff}$=1.

The trimer excitations are more complex and require a second labelling to describe the levels corresponding to $\vec{S}_{13}=\vec{S}_{1}+\vec{S}_{3}$.  Only the lowest energy excitations which having dipole allowed and non zero neutron cross sections were included in the fit in the main text.

\textit{Open data access:} Following UK research council guidance, data files can be accessed either at source (from the ILL (www.ill.eu), or the NCNR (www.ncnr.nist.gov) or through the University of Edinburgh's online digital repository (datashare.is.ed.ac.uk) after publication.  

\bibliography{CFO_bib}